\documentclass{article}
\usepackage{frascatiphys,here,graphicx,subfigure}

\def \de {\Delta E}
\def \kkp {D^+\to K^+ K^-\pi^+}

\begin{document}
\setlength{\tabcolsep}{4pt}
\title{DALITZ PLOT ANALYSES AT CLEO-C}
\author{Paras Naik      \\
{\em Carleton University, Ottawa, Ontario, Canada K1S 5B6} \\
Liming Zhang   \\
{\em Syracuse University, Syracuse, New York, USA 13244}\\
Norman Lowrey       \\
{\em University of Illinois at Urbana-Champaign, Urbana, Illinois,
USA 61801}\\
CLEO Collaboration   \\
{\em LEPP, Ithaca, New York, USA 14853} }
\maketitle
\baselineskip=11.6pt
\begin{abstract}
We present several recent analyses of Dalitz plots from the CLEO-c
experiment, including published and preliminary analyses of $D^+ \to
\pi^- \pi^+ \pi^+$, $D^+ \to K^- \pi^+ \pi^+$, and $D^0 \to
K^0_{S,L} \pi^+ \pi^-$ decays. More information on these analyses
can be found in References \cite{Mikhail1, Mikhail2, Eric}. New
preliminary analyses we present include a search for $CP$ asymmetry
in $D^+ \to K^+ K^- \pi^+$ decays and a Dalitz plot analysis of $D^0
\to K^0_{S} \pi^0 \pi^0$.

We report on a search for the $CP$ asymmetry in the singly
Cabibbo-suppressed decay ${D^+\to K^+ K^-\pi^+}$ using a data sample
of 572 pb$^{-1}$ accumulated with the CLEO-c detector and taken at
the $e^+ e^- \to \psi(3770)$ resonance. We have searched for $CP$
asymmetries using a Dalitz plot based analysis that determines the
amplitudes and relative phases of the intermediate states.

We also use a 281 pb$^{-1}$ CLEO-c data sample taken at the $e^+ e^-
\to \psi$(3770) resonance to study the $D^0 \rightarrow K^0_S \pi^0
\pi^0$ Dalitz plot. Our nominal fit includes the $K^0_S$,
$K^*(892)$, $f_0(980)$, $f_0(1370)$, and $K^*(1680)$ resonances.
\end{abstract}

\baselineskip=14pt

\section{Search for $CP$ asymmetry in $D^+ \to K^+ K^- \pi^+$ Decays}
Singly Cabibbo-suppressed (SCS) $D$-meson decays are predicted in
the Standard Model (SM) to exhibit $CP$-violating charge asymmetries
smaller than the order of $10^{-3}$. Direct $CP$ violation in SCS
decays could arise from the interference between tree-level and
penguin processes. Doubly Cabibbo-suppressed and Cabibbo-favored
(CF) decays are expected to be $CP$ invariant in the SM due to the
lack of contribution from penguin processes. Measurements of $CP$
asymmetries in SCS processes greater than $\mathcal {O}(10^{-3})$
would be evidence of physics beyond the SM \cite{nir}.

\begin{figure}[h]
\center
\includegraphics[width=0.70\textwidth]{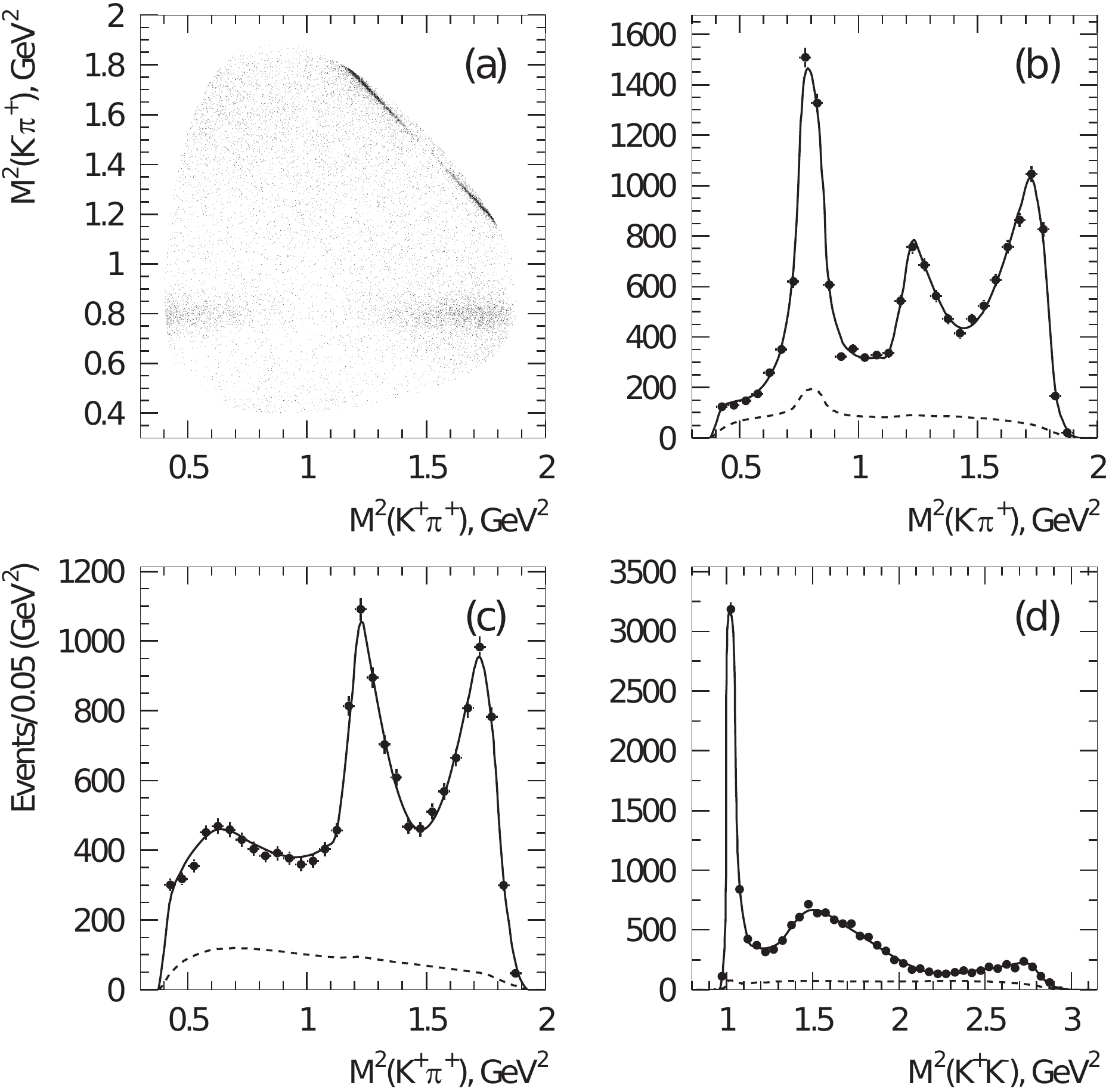}
\caption{\it \label{fig:lass}The results of fitting the $\kkp$ data
for Model three. (a) The scatter plot for squared mass of $K^-\pi^+$
versus $K^+\pi^+$ and the projections onto squared mass of (b)
$K^-\pi^+$ (c) $K^+\pi^+$ and (d) $K^+K^-$ for both fit (curve) and
data (points) are shown. The dashed line shows the background
contribution. }
\end{figure}

We define two variables: the energy difference $ \Delta E \equiv
\sum_iE_i-E_{\rm beam}$ and the beam-constrained mass $ m_{\rm BC}
\equiv \sqrt{E_{\rm beam}^2-|\sum_i \vec{P_i}|^2}, $ where $E_i$,
$\vec{P_i}$ are the energy and momentum of each $D$ decay product,
and $E_{\rm beam}$ is the beam energy. We define a signal box
corresponding to 2.5 standard deviations in each variable, and
remove multiple candidates in each event by choosing the candidate
that gives the smallest $|\de|$. We obtain $13693\pm137$ $\kkp$
signal candidates. To reduce smearing effects introduced by the
detector, a mass constraint fit for the $D^+$ candidate is applied
to obtain the mass squared variables, $m^2_{K^+\pi^+}$ and
$m^2_{K^-\pi^+}$, for the $D^+\to K^+ K^-\pi^+$ Dalitz plot (DP)
shown in Figure \ref{fig:lass}(a).

The decay amplitude as a function of DP variables is expressed as a
sum of two-body matrix elements and one non-resonant (NR) decay
amplitude \cite{kopp}. For most resonances, the matrix element is
parameterized by Breit-Wigner shapes that take into account $D$
meson and intermediate resonance form factors and angular
dependence. For the $f_0(980)$ we use a Flatt\'e function \cite{f0}.
For the $a_0(980)$, we use the function in Ref. \cite{a0}. We choose
the same phase conventions for the intermediate resonances as the
E687 Collaboration \cite{e687}. A fit fraction (FF), the integral of
a single component divided by the sum of all components, is reported
for each intermediate resonance to allow for more meaningful
comparisons between results.

For $D^+$ decays to $K^-\pi^+$ $S$-wave states, we consider three
amplitude models. One model uses a coherent sum of a uniform
non-resonant term and Breit-Wigner term for the $K_0^*(1430)$
resonance. The second model only uses a Breit-Wigner term for the
$K_0^*(1430)$ resonance. The third model uses the LASS amplitude for
$K^-\pi^+\to K^-\pi^+$ elastic scattering \cite{LASS,BABARLASS}. We
present results only for the third model, although the first model
provides a similar fit.

\begin{table}[h]
\center \caption{\it \label{tab:lass} The fit results in Model
three. The errors shown are statistical, experimental systematic,
and modeling systematic respectively.}
\begin{tabular}{rccc}\hline
Component &Amplitude &Phase ($^\circ$)&Fit Fraction (\%)\\\hline
$\overline{K}^*(892)^0K^+$& 1(fixed)&0(fixed)&$23.9\pm0.6 ^{+0.1+0.9}_{-0.3-0.4}$\\

$K^-\pi^+(S)K^+$&$4.53\pm0.16^{+0.22+0.31}_{-0.01-0.23}$&$21\pm3^{+0+7}_{-6-2}$&
$53\pm3^{+5+8}_{-0-5}$ \\

$a_0(980)\pi^+$&$0.74\pm0.09^{+0.03+0.16}_{-0.01-0.39}$&$96\pm7^{+0+4}_{-4-15}$&
$1.7\pm0.4^{+0.1+1.3}_{-0.0-0.6}$\\

$\phi(1020)\pi^+$&$1.23\pm0.02^{+0.00+0.01}_{-0.00-0.02}$&$-148\pm3^{+1+5}_{-1-3}$&
$28.0\pm0.5^{+0.0}_{-0.4}\pm0.5$\\

$f_2(1270)\pi^+$&$0.91\pm0.13_{-0.01-0.24}^{+0.03+0.11}$&$20\pm6^{+5+9}_{-0-11}$&
$0.9\pm0.2^{+0.1}_{-0.0}\pm0.2$\\

$a_0(1450)\pi^+$&$1.36\pm0.10^{+0.20+0.45}_{-0.01-0.25}$&$116\pm5^{+1+13}_{-5-10}$&
$3.4\pm0.5^{+1.0+2.5}_{-0.0-1.2}$\\

$\phi(1680)\pi^+$&$2.6\pm0.3^{+0.2+0.6}_{-0.0-0.7}$&$-96\pm10^{+0+17}_{-16-12}$&
$0.89\pm0.18^{+0.15+0.3}_{-0.02-0.2}$\\

$\overline{K}_2^*(1430)^0K^+$&$3.5\pm1.0^{+1.6+1.6}_{-0.0-2.6}$&$-156\pm6_{-0-8}^{+1+30}$&
$2.1\pm1.2^{+2.4+2.2}_{-0.0-1.3}$\\

 \hline
\end{tabular}
\end{table}
We determine the detection efficiency as a function of the two DP
variables by fitting a signal MC sample generated with a flat
distribution in the phase space. We use a fit to the events in the
$\Delta E$ sideband ($24<|\Delta E|<42$ MeV and $|m_{\rm
BC}-m_{D^+}|<9$ MeV/$c^2$) to describe the background distribution
of the DP. Having information for both the background and
efficiency, as well as the fraction of signal events in the signal
region, we fit the data in the DP to extract the amplitudes and
phases of any contributing intermediate resonances. We perform an
unbinned maximum likelihood fit. The signal fraction $f$ is
$f_0=(84.1\pm0.2$)\%, constrained in the fit to be within its error
$\sigma_f$ obtained from the fit to the $m_{\rm BC}$ distribution.
We begin by fitting the DP with all known resonances that may
possibly contribute to this decay. We determine which resonances are
to be included by maximizing the fit confidence level (C.L.). The
procedure is to add all possible resonances, then subsequently
remove those which do not contribute significantly, or worsen our
C.L. The projections of the DP for the fit to Model three are shown
in Figures \ref{fig:lass}(b-d). The results of the fit amplitudes,
phases, and fractions including errors are shown in Table
\ref{tab:lass} for Model three.

\begin{table}[h]
\center
 \caption{\it \label{acp:lass}
$A_{CP}$ for each component of the fit
 using $D^\pm$
samples in Model three. The errors for fit fractions and phases are
statistical only, and those for $A_{CP}$ are statistical,
experimental systematic, and modeling systematic respectively.}
\begin{tabular}{r|c}\hline
Component $j$ & ${A_{CP}}_j$(\%)\\\hline
$\overline{K}^*(892)^0K^+$ & $-0.1\pm2.9_{-0.4-0.4}^{+2.3+0.7}$\\
$K^-\pi^+(S)K^+$&$-1\pm5_{-2-4}^{+1+6}$\\
$a_0(980)\pi^+$&$-11\pm23_{-9-6}^{+4+24}$\\
$\phi(1020)\pi^+$&$-3.0\pm1.9_{-0.2-0.3}^{+0.1+0.2}$\\
$f_2(1270)\pi^+$&$4\pm25_{-4-46}^{+3+22}$\\
$a_0(1450)\pi^+$&$-18\pm14_{-8-9}^{+0+16}$\\
$\phi(1680)\pi^+$&$-9\pm21_{-4-3}^{+22+7}$\\
$\overline{K}_2^*(1430)^0K^+$&$69\pm51_{-28-41}^{+1+8}$\\
\hline
\end{tabular}
\end{table}
To search for $CP$ violation in this model, we fit the $D^+$ and
$D^-$ samples independently. We use the same background fraction and
PDF as those used in the fit to the total sample, but different
coefficients for efficiency functions which are obtained from signal
MC of $D^\pm$ decays. The calculated $CP$ asymmetry, ${A_{CP}}_j
\equiv \frac{{FF_j}_{D^+}-{FF_j}_{D^-}}{{FF_j}_{D^+}+{FF_j}_{D^-}}$,
is shown for each resonance $j$ in Table \ref{acp:lass}.

\section{Dalitz Plot Analysis of $D^0 \to K^0_S \pi^0 \pi^0$ Decays}

The PDG \cite{PDG} has little information on the $D^0 \rightarrow
K^0_S \pi^0 \pi^0$ decay. In addition to providing a more
comprehensive study of the $D^0 \rightarrow K^0_S \pi^0 \pi^0$
decay, this DP analysis seems like a good place to look for the low
mass $\pi \pi$ $S$-wave signature of the $\sigma$. The $K^0_S \pi^+
\pi^-$ mode is much cleaner and has better statistics, but the
$\rho^0$ resonance overlaps the region where we would expect to find
the low mass $S$-wave signature. Using CLEO-c data, we eliminate
nearly all of the background by doing a double-tagged analysis,
where both $D$ mesons are completely reconstructed.

We have analyzed $281$ pb$^{-1}$ of CLEO-c data taken on the $e^+e^-
\to \psi(3770)$ resonance. In a double-tagged analysis, both $D$
mesons are reconstructed.  For our double-tagged analysis, we
consider candidates with one $D$ reconstructed as $K^0_S\pi^0\pi^0$,
and the other $D$ reconstructed using any of the following decay
modes (charge conjugation is implied throughout this analysis):
$\overline{D^0} \rightarrow K^+\pi^-$, $\overline{D^0} \rightarrow
K^+\pi^- \pi^0$, $\overline{D^0} \rightarrow K^+\pi^-\pi^+\pi^-$. In
a single-tagged analysis, we reconstruct only one $D$ meson in the
event, which decays to $K^0_S\pi^0\pi^0$.

\begin{table}[h]
  \begin{center}
    \caption{\it $D^0 \rightarrow K^0_S\pi^0\pi^0$ signal yield, number of candidates, and signal fraction}
    \label{t_signalfraction}
    \begin{tabular}{|c||c|c|}  \hline
      Result           & Double Tag        &  Single Tag         \\ \hline
      Signal Yield     & 257 $\pm$ 17      &  1884 $\pm$ 56      \\
      Total Candidates & 276               &  2548               \\
      Signal Fraction  & 0.931 $\pm$ 0.062 &  0.739 $\pm$ 0.022  \\ \hline
    \end{tabular}
  \end{center}
\end{table}
To reduce $2\pi$ background that fakes a $K^0_S$, we enforce a
$2\sigma$ enhanced flight significance selection criteria on our
$K^0_S$ candidates. To reduce the $K\pi\pi^0$ background, we require
$|dE/dx_{pion}|$ $<$ $3\sigma$ and $dE/dx_{kaon}$ $<$ $-2\sigma$ for
both  $K^0_S$ daughter pions. We use the same particle
identification selection criteria for double-tagged and
single-tagged analyses. We apply a $2\sigma$ selection criteria on
the reconstructed $K^0_S$ mass. After enforcing our selection
criteria on the $K^0_S$ mass, we apply a $2\sigma$ selection
criteria on $\Delta E$. We additionally apply a $2\sigma$ cut on the
beam constrained mass. For each event that has more than one
candidate, we require the following: For the double-tagged data, we
take the average of the signal beam constrained mass and the tagged
beam constrained mass, and we select whichever candidate's average
is closest to the nominal $D$ mass. For the single-tagged data, we
select the candidate with $\Delta E$ closest to zero. Table
\ref{t_signalfraction} shows our signal yield and signal fraction.

For this analysis, we define our DP variables as follows: $x$
$\equiv$ larger $m_{K^0_S\pi^0}^2$, $y$ $\equiv$ $m_{\pi^0\pi^0}^2$,
$z$ $\equiv$ smaller $m_{K^0_S\pi^0}^2$. When fitting such a Dalitz
plot, we must take into account the fact that the two $\pi^0$ final
state particles are indistinguishable, so we explicitly symmetrize
the functions we use in $x$ and $z$.

To study the efficiency of reconstructing our signal, we generate
100000 signal Monte Carlo events distributed uniformly across the
Dalitz plot phase space.  Half of these events force the $D^0$ to
decay directly into $K^0_S\pi^0\pi^0$ and the $\overline{D}^0$ to
decay into neutrinos.  The other half of these events force the
$\overline{D}^0$ to decay directly into our signal mode and the
$D^0$ to decay into neutrinos. We fit the efficiency over the Dalitz
plot to a third-order polynomial explicitly symmetric in $x$ and
$z$. To fit for the background, we use a sideband from single-tagged
data which is centered 5$\sigma_{m_{D^0}}$ lower in $m_{BC}$ than
the signal region, with the same width as that of the signal region,
and has the appropriate range in $\Delta{E}$ which conserves the
boundaries of the signal DP. We use this background shape for the
double-tagged data as well as for the single-tagged data. We fit the
background events to a third-order polynomial explicitly symmetric
in $x$ and $z$, plus a non-interfering $K^*(892)$ Breit-Wigner in
both $x$ and $z$.

The signal is parameterized with an isobar model that has four
interfering resonances plus one non-interfering resonance. To
enforce the symmetry requirement in the DP, we include each $K^*$
resonance as an $x$ resonance and a $z$ resonance, while using the
same amplitude and phase for the $x$ contribution and $z$
contribution. The parameters for the $K^0_S$, $K^*(892)$, and
$K^*(1680)$ come from the PDG \cite{PDG}.  The parameters for the
$f_0(980)$ are approximated from a BES paper \cite{BES}. The
parameters for the $f_0(1370)$ come from Reference \cite{f1370}.

\begin{figure}[h]
\begin{center}
        {\includegraphics[width=0.33733218634197988353626257278984\textwidth]{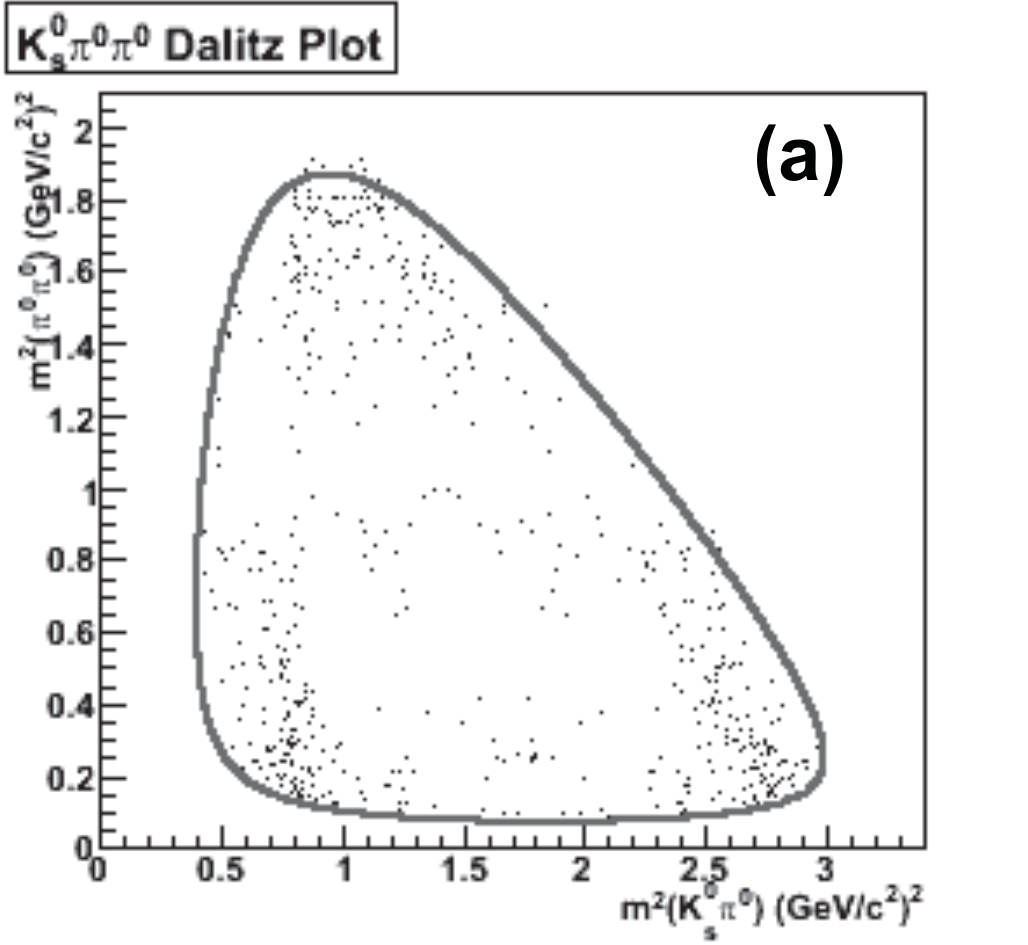}}
        {\includegraphics[width=0.31\textwidth]{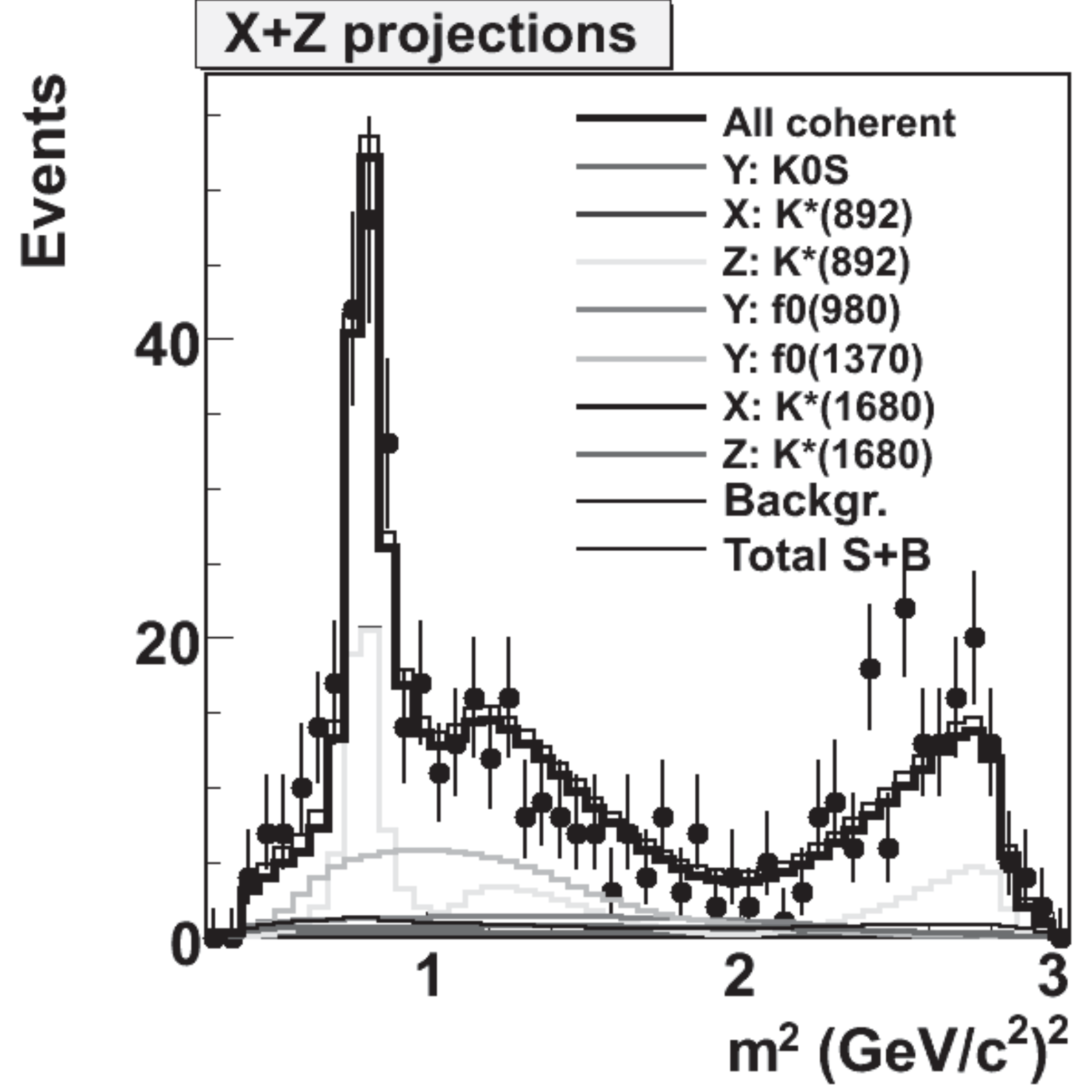}}
        {\includegraphics[width=0.31\textwidth]{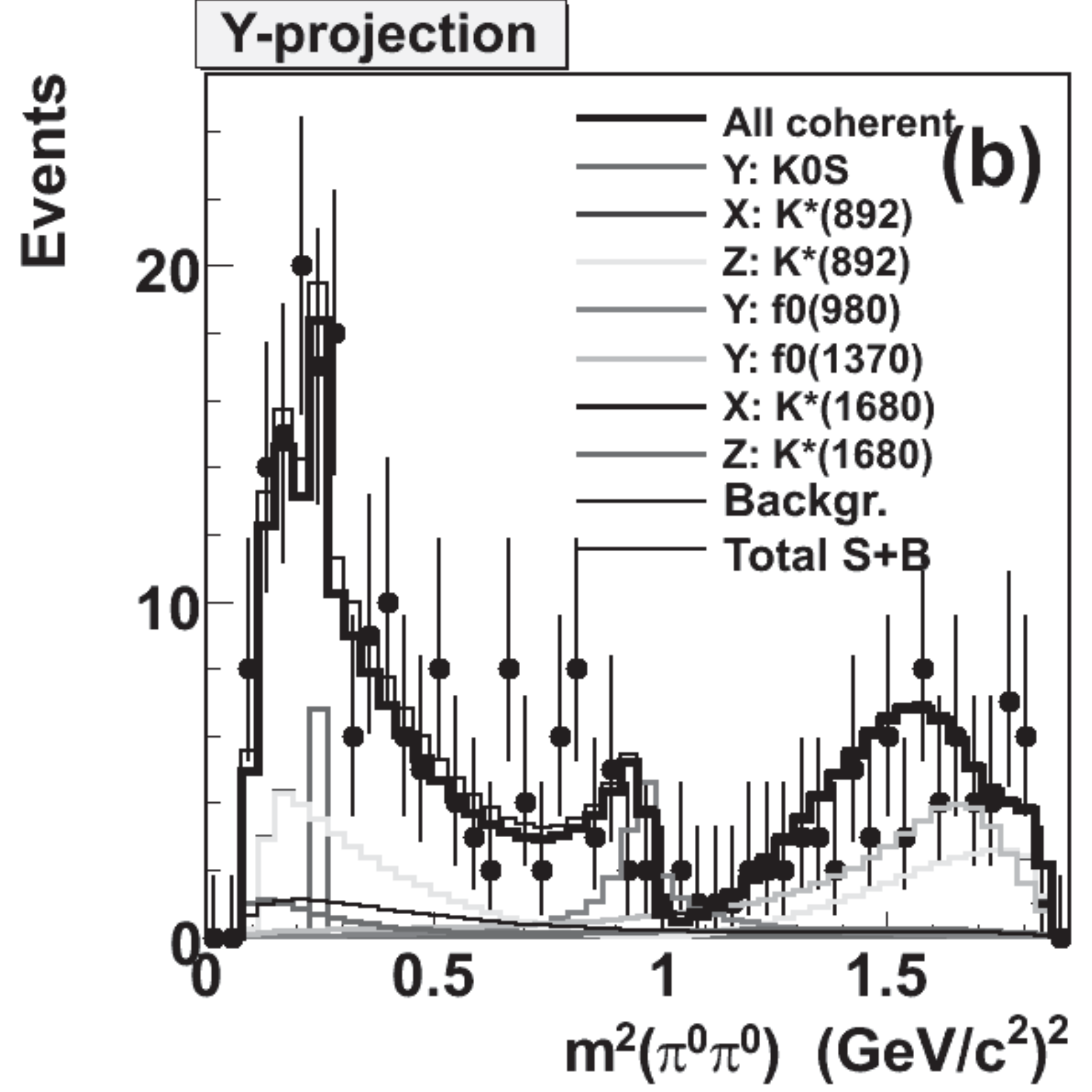}}
        \caption{\it \label{f_fit_dbl_all}(a) Dalitz plot of the double-tagged data (2 entries per candidate, second entry has $x$ and $z$ swapped) and (b) fits to the double-tagged x+z projection and double-tagged y projection.}
\end{center}
\end{figure}
Figure \ref{f_fit_dbl_all}(a) displays the DP from the double-tagged
data. To fit this DP with an unbinned maximum likelihood fitter, we
fix the signal fraction to 0.931 as determined from the beam
constrained mass distribution. The fit also fixes the efficiency
parameters and background parameters as determined from the signal
Monte Carlo and sideband. The fit determines the amplitudes and
phases of the resonances and calculates the fit fractions. Figure
\ref{f_fit_dbl_all}(b) shows the fit results.

\begin{table}[h]
  \begin{center}
    \caption{\it Final results including systematic errors.  The first error is statistical.  The second error is systematic arising from our selection criteria.  The third error is systematic arising from our signal model.}
    \label{t_finalResults}
    \begin{tabular}{|c||c|c|}  \hline
      resonance & & double tag  \\ \hline

      &Fit Fraction & $0.026 \pm 0.023 \pm 0.003 \pm 0.001$      \\
      $K^0_S$ & Amplitude    & $0.101 \pm 0.029 \pm 0.009 \pm 0.004$      \\
      &Effective Width   & $0.0046 \pm 0.0011 \pm 0.0001 \pm 0.0001$  \\ \hline

      &Fit Fraction & $0.542 \pm 0.054 \pm 0.030 \pm 0.053$  \\
      $K^*(892)$ &Amplitude    & $1$ (fixed)                            \\
      &Phase ($^\circ$)  & $0$ (fixed)                            \\ \hline

      &Fit Fraction & $0.090 \pm 0.032 \pm 0.009 \pm 0.027$  \\
      $f_0(980)$&Amplitude    & $1.50 \pm 0.27 \pm 0.10 \pm 0.20$      \\
      &Phase ($^\circ$)       & $12 \pm 17 \pm 14 \pm 8$               \\ \hline

      &Fit Fraction & $0.238 \pm 0.071 \pm 0.047 \pm 0.086$  \\
      $f_0(1370)$ &Amplitude    & $2.77 \pm 0.45 \pm 0.30 \pm 0.66$      \\
      &Phase ($^\circ$)       & $344 \pm 10 \pm 10 \pm 18$             \\ \hline

      &Fit Fraction & $0.114 \pm 0.027 \pm 0.021 \pm 0.032$  \\
      $K^*(1680)$&Amplitude    & $4.55 \pm 0.68 \pm 0.49 \pm 0.59$      \\
      &Phase ($^\circ$)       & $97 \pm 20 \pm 17 \pm 13$              \\ \hline
    \end{tabular}
  \end{center}
\end{table}
To estimate systematic errors, we use the technique developed by Jim
Wiss and Rob Gardner \cite{EstSys}.  Using this technique, the
systematic errors are essentially independent of the number of
systematic sources considered \cite{EstSys}. Table
\ref{t_finalResults} gives our preliminary results. We are currently
extending our analysis to the full available CLEO-c $\psi(3770)$
data sample, and studying the effects of using a $\sigma$ or
$\kappa$ $S$-wave to possibly improve our fit.

\section{Acknowledgements}
We would like to thank David Asner, David Cinabro, Mikhail Dubrovin,
Qing He, Mats Selen, Ed Thorndike, Eric White, our paper committees,
and the rest of the CLEO-c Dalitz Plot Analysis Working Group for
their insight, comments, and suggestions.

\end{document}